\documentclass[aps,prl,twocolumn,superscriptaddress,footinbib,longbibliography]{revtex4}
\usepackage{bm}

\usepackage{graphicx}
\usepackage{amsmath}
\usepackage{hyperref}
\usepackage{natbib}
\usepackage{xcolor,colortbl}

\newcommand {\rmi}{{\rm i}}
\newcommand {\rmd}{{\rm d}}

\renewcommand {\phi}{\varphi}
\renewcommand {\epsilon}{\varepsilon}
\newcommand {\eps}{\varepsilon}
\newcommand{\nix}[1]{}

\renewcommand {\Im}{\mathop{\mathrm{Im}}}

\begin{document}
\author{Andrey A. Stashkevich}
\affiliation{LSPM (CNRS-UPR 3407), Universit\'e Paris 13, Sorbonne Paris  Cit\'e,  93430 Villetaneuse, France}
\affiliation{International laboratory ``MultiferrLab'', ITMO University, St.~Petersburg 197101, Russia}
\author{Yves Roussign\'e}
\affiliation{LSPM (CNRS-UPR 3407), Universit\'e Paris 13, Sorbonne Paris  Cit\'e,  93430 Villetaneuse, France}
\author{Alexander~N. Poddubny}
\affiliation{ITMO University, St.~Petersburg 197101, Russia}
\affiliation{Ioffe Institute, St. Petersburg 194021, Russia}
\email{a.poddubny@phoi.ifmo.ru}
\author{ S.-M. Ch\'erif }
\affiliation{LSPM (CNRS-UPR 3407), Universit\'e Paris 13, Sorbonne Paris  Cit\'e,  93430 Villetaneuse, France}
\author{Y. Zheng}
\author{Franck Vidal}
\affiliation{Institut des NanoSciences de Paris, UMR CNRS 7588, UPMC Universit\'e Paris 6, 4 Place Jussieu 75005 Paris, France}
\author{Ilya~V. Yagupov}
\author{Alexei~P. Slobozhanyuk}
\author{Pavel~A. Belov}
\affiliation{ITMO University, St.~Petersburg 197101, Russia}
\author{Yuri~S. Kivshar}
\affiliation{ITMO University, St.~Petersburg 197101, Russia}
\affiliation{Nonlinear Physics Center, Australian National University, Canberra ACT 0200, Australia}

\title{
Anomalous polarization conversion in arrays of ultrathin ferromagnetic nanowires}

\begin{abstract}
We study optical properties of arrays of ultrathin nanowires by means of the Brillouin scattering of light on magnons.
We employ the Stokes/anti-Stokes scattering asymmetry to probe the circular polarization of a local electric field induced inside
nanowires by linearly polarized light waves. We observe the anomalous polarization conversion of the opposite sign than that
in a bulk medium or thick nanowires with a great enhancement of the degree of circular polarization attributed
to an unconventional refraction in the nanowire medium.
\end{abstract}

\maketitle

The study of magneto-optical response of tailored nanostructures is in the focus of active research of nanostructured materials~\cite{Belotelov2013,Chin2013,Luca2014}. Nonmagnetic metallic nanowires are well known in optics, and they are employed as building blocks of the so-called {\em wire metamaterials}~\cite{simovski2012}. Such structures demonstrate many unusual properties, including negative refraction~\cite{Yao2008}, enhanced sensing~\cite{kabashin2009}, super-lensing~\cite{Fink2012}, strong nonlocal effects~\cite{belov2003}, nonlinearities~\cite{Ginzburg2013}, and they can boost light-matter interaction in the regime of the hyperbolic dispersion~\cite{review2013}.

\begin{figure}[b]
\includegraphics[width=0.55\textwidth]{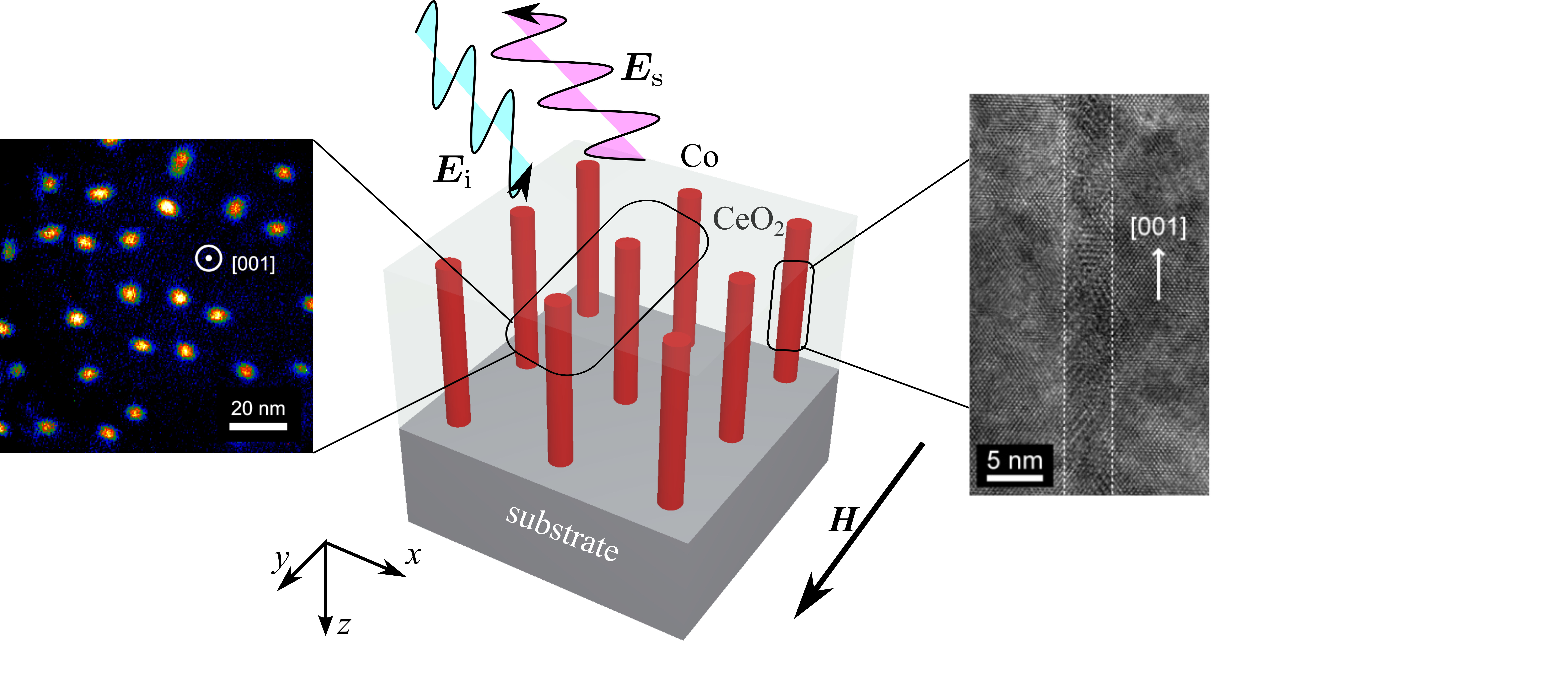}
\caption{Schematics of the Co nanowires and the Brillouin scattering geometry.
Incident ($p$-polarized) and scattered ($s$-polarized) photons and the direction of the magnetic field $\bm H$ are indicated.
Left inset: TEM image, acquired in energy-filtered mode at the Co L edge, taken in plane-view geometry. Right inset: HRTEM image of a single nanowire taken in cross-section.
 }\label{fig:1}
\end{figure}

Here, we study magneto-optical (MO) properties of ultrathin ferromagnetic Co nanowire arrays (see Fig.~\ref{fig:1}) by means of polarization-resolved Brillouin light scattering on magnons~\cite{CardonaVII,Wright1969,Cottam1986}. In a sharp contrast to the previous studies of ferromagnetic nanowires with the diameter $D$ of 20~nm \cite{Stashkevich2009}, here we study arrays of thin nanowires with diameter $D\sim 4.8$~nm and lattice spacing $\sim 17$~nm (Fig.~\ref{fig:1}). We reveal a striking anomaly of Stokes/anti-Stokes pattern in the spectra of Brillouin Light Scattering (BLS) from magnons in such structures. First, its asymmetry is inverted, which is impossible in principle in continuous non-structured metal layers; secondly, it is unconventionally large. Although the latter effect can be predicted formally for continuous films, such a huge asymmetry can only be calculated for materials with unrealistic optical parameters. This is explained by a strong  modification of  optical and magneto-optical properties for thin nanowires. While  the coherent propagation of photons in the sample is weakly affected by thin wires due to their small volume concentration, the photon interaction with magnons is confined to the volume inside the wires where the wave polarization is strongly modified due to a huge mismatch in the dielectric constants of metallic nanowires and dielectric media. As a result, we observe that a linearly polarized light obliquely incident upon a metacrystal composed of ultrathin Co nanowires acquires ellipticity inside the wires that is enhanced by an order of magnitude being of the opposite sign as that in the case of continuous Co films or thicker Co wires. We visualize this effect by measuring the asymmetry of Stokes and anti-Stokes peaks in the Brillouin scattering spectra of light by the spin-wave modes of the wires, since the scattering can probe local electric fields~\cite{Dahl1995}. We expect that our results will be instrumental for the emerging field of  nonlinear spectroscopy of metamaterials~\cite{our_thg} as well as for a design of novel structures with strong chiral responses~\cite{Cohen2010} and  polarization-sensitive light rooting~\cite{Zayats2013,Kapitanova2014,Petersen2014}.

An array of ultrathin Co nanowires  has been grown by  sequential pulsed laser deposition  of Co and CeO$_{2}$ in reductive conditions ($P_{\rm growth} = 10^{-5}$~mbar), leading to self-assembly of Co nanowires embedded in a CeO$_{2}$ matrix on a  SrTiO$_{3}$(001) (SurfaceNet GmbH) substrate using a quadrupled Nd:YAG laser (wavelength 266 nm) operating at 10 Hz and a fluence in the 1-3 J$\times$cm$^{-2}$ range. More details are given in Refs.~\cite{Vidal2012,Bonilla2013}. Metallic nanowire formation in the sample was evidenced using high resolution and energy-filtered transmission electron microscopy data (acquired using a JEOL JEM 2100F equipped with a field-emission gun operated at 200 kV and a Gatan GIF spectrometer), see the right inset of Fig.~\ref{fig:1}. The wires are oriented perpendicular to the surface of the substrate and extend throughout the matrix thickness, $t$. Importantly, the technology employed is such that the wire length $h$ coincides with the film thickness $t$, i.e. $t = h$. Their length turned out to be equal to 470 nm. The diameter, $D=2R=4.8\pm 0.7$~nm of the wires was determined by collecting images in a planar geometry.

Now we proceed with the analysis of the spectra of Brillouin Light Scattering~(BLS) from thermal magnons localized on ferromagnetic nanowires. The experimental arrangement is sketched in Fig.~\ref{fig:1}, and it corresponds to the Damon-Eshbach geometry~\cite{Damon1961}. A magnetic field $\bm H$ is applied in the plane of the sample. The plane of incidence is perpendicular to the applied field. The incident laser beam is $p$-polarized with wavelength $\lambda_{\rm opt}=532~$nm. The backscattered light is probed in $s$-polarization through a tandem Fabry Perot interferometer (JR Sandercock product).
The BLS process is a particular case of nonlinear wave-mixing, and it generates at the output two frequency shifted optical waves, namely a down-shifted, known both in Raman and Brillouin spectroscopy as the Stokes (S) line, and up-shifted called the anti-Stokes (AS) line.
Typically, light scattering spectra are asymmetric, i.e. the amplitudes of S and AS spectral lines are not equal. However, the physical mechanisms producing this peculiar asymmetry are completely different. In the Raman case it is due to a greater difference between the frequencies of S and AS lines which results in an appreciable asymmetry in the density of states corresponding to the frequencies $\omega + \Omega$ and $\omega - \Omega$, where $\omega$ and $\Omega$ are incoming photon and magnon frequencies, respectively.
 In the BLS process the frequency shifts are smaller by several orders of magnitude, and entirely different physical effects are involved,  namely a very particular symmetry of MO interactions. Mathematically, the symmetry of MO coupling   is described by the totally antisymmetric Levi-Civita tensor~\cite{Cottam1986,CardonaVII}. As a result, the efficiency of the MO interactions is expressed via the mixed product $\bm E _{\rm s}\cdot (\bm m\times\bm E_{\rm i})$ of the polarizations of the interacting waves: the incident ($\bm E_{i}$) and scattered ($\bm E_{s}$) optical waves  and the scattering spin wave ($\bm m$).  Importantly, in the general case of complex vector space this mixed product is not invariant with respect to complex conjugation of the waves. In physical terms it is the elliptical polarization that is linked  to the complex vectors and the complex conjugation corresponds to the inversion of the direction of rotation of such polarization. This crucially important point will be revisited below.

\begin{figure}[t!]
\centering\includegraphics[width=0.8\columnwidth]{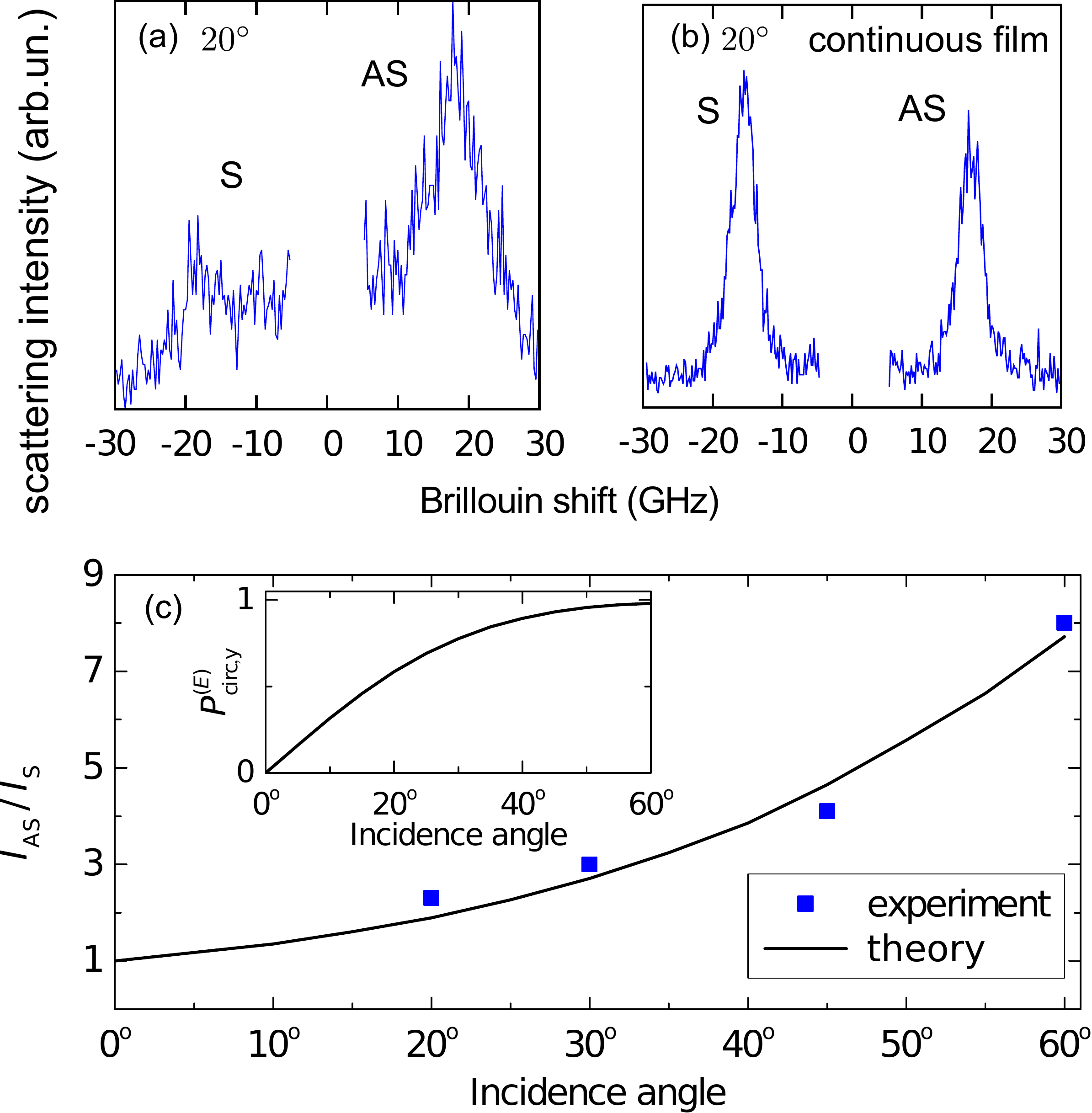}
\caption{(a) Experimental Brillouin scattering spectrum of the nanowire array for
 three incidence angles $\theta=20^{\circ}$ (a).
 with an  applied in-plane magnetic field of 7~kOe. (b) Spectrum for the 1.2 nm thick  continuous Co film  with the field of 6~kOe.
 (c) Measured (squares) and calculated (line)  dependence of the ratio of Stokes and anti-Stokes spectral peak heights in the incidence angle $\theta$. The inset shows the circular polarization degree of the electric field inside the wires at $y=x=0$, averaged over the wire length. The calculation details are given in text.
 }\label{fig:spectra}
\end{figure}

Experimental BLS spectra for the angle of incidence $\theta=20^{\circ}$ are presented in the panel (a) of Fig.~\ref{fig:spectra}. The spectra are not symmetric: the down-shifted Stokes line and the up-shifted anti-Stokes  line have different magnitudes, which is not untypical of magneto-optical BLS spectra~\cite{Stashkevich2009}. What is really not conventional, however, is an extremely high degree of the Stokes/anti-Stokes (S/AS) asymmetry and even more so its inversion with respect to its ``classical'' pattern in which the intensity of the down-shifted Stokes peak $|E_S |^2$ is greater than that of the up-shifted anti-Stokes one $|E_{AS} |^2$, i.e. $|E_{S} |^2>|E_{AS} |^2$. The latter is illustrated in Fig.~\ref{fig:spectra}(b). This dramatic reversal of the scattering spectra asymmetry constitutes the main result of our work and is analyzed in detail below.

Now we proceed to the theoretical analysis of the Stokes/anti-Stokes asymmetry in the scattering spectra. In this respect, identification of the spin wave (SW) modes contributing to the BLS spectra is very important.
The applied static magnetic  field of 7 kOe fully saturates the sample so that magnetization in each wire is perfectly homogeneous. The latter allows a reliable theoretical description of the spin-wave behavior, including explicit expressions for magnetization.

The only candidate for the role of the ``effectively scattering mode'' are the vertical Kittel SW modes, i.e. spin waves propagating along the wire axis with the wave number $K$ with uniform cross-section distribution (at least, in the approximation  $KR\ll 1$).
The limiting case of $K = 0$ corresponds to a purely magnetostatic perfectly uniform oscillations~\cite{KittelIntro},  and it is characterized by the frequency  $(\Omega/\gamma)^{2}=H(H-2\pi M_{\rm eff})$. Here,  $M_{\rm eff}$ takes into account the dipolar inter-wire interactions between static magnetizations in individual nanowires, which makes it slightly smaller than the conventional bulk value for the cobalt, and $\gamma$ is the gyromagnetic ratio.
The state of ellipticity of magnon polarization is crucial in estimating the S/AS asymmetry. Its actual value is a result of a trade-off between two trends, namely, a purely circular shape imposed by the ferromagnetic resonance and the flattening effect of dipolar interactions. Supplemental Material includes more details on  the dispersion relation $\Omega(K)$, spatial distribution and ellipticity of the spin waves.

\begin{figure}[t!]
\centering\includegraphics[width=0.45\textwidth]{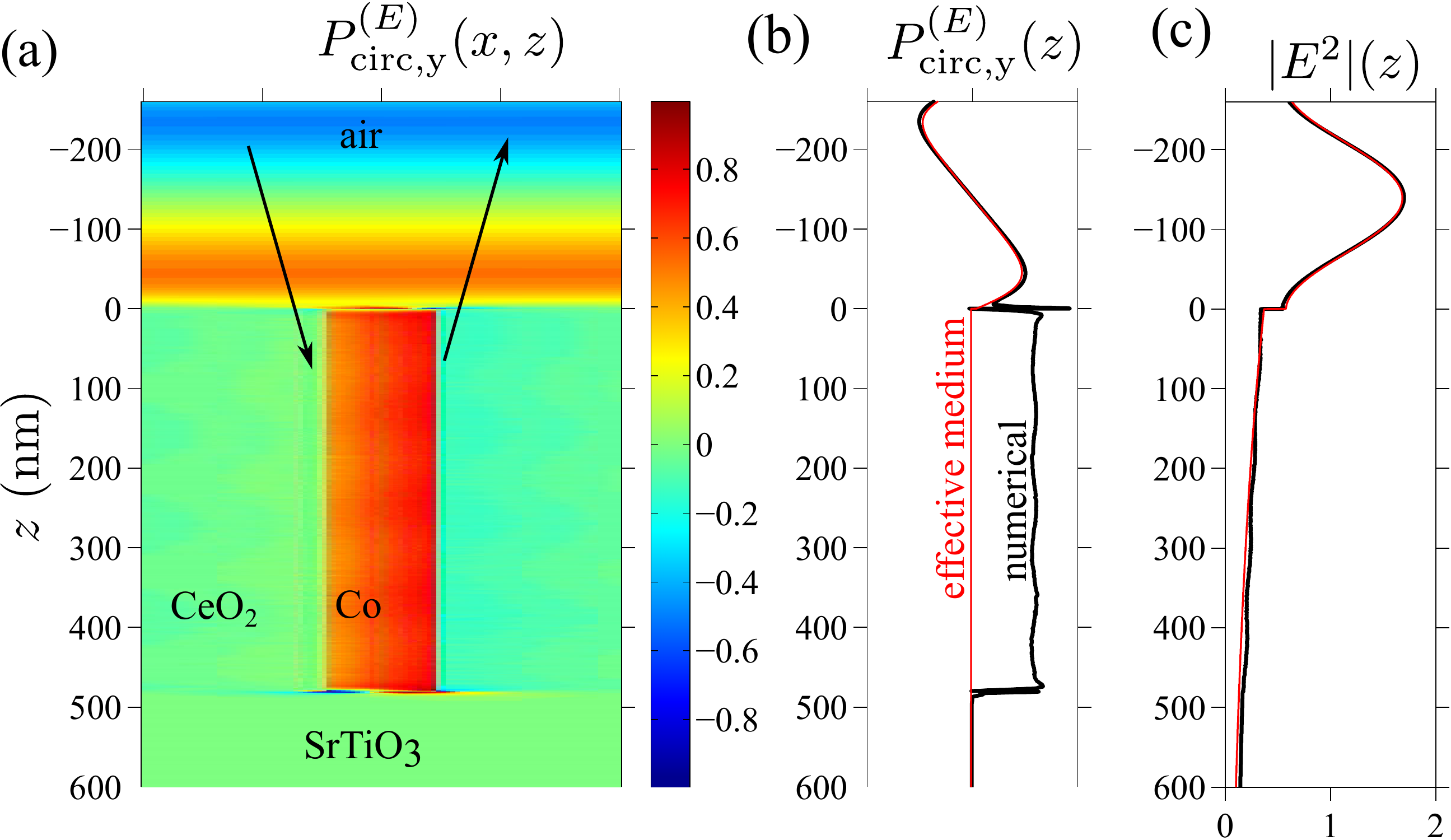}
\caption{(a) Theoretically calculated spatial maps of the  circular polarization of the electric field in the unit cell of Co nanowire array. The arrows show the propagation directions of incident and scattered waves. (b) The cross-sections of the spatial map along the wire centers $x=y=0$, calculated numerically (black/solid lines) and analytically (red/thin lines). (c)  Intensity of the electric field averaged over $x$ coordinate within the unit cell at $y=0$,.
 The calculation has been performed for  $\lambda_{\rm opt}=532~$nm, $\theta=20^{\circ}$,  TM polarization, $d=4.8$~nm, $a=20$~nm,
  $\eps_{\rm Co}=-10.4+17.1\rmi$~\cite{Johnson1974},  $\eps_{\rm CeO2}=4.8$~\cite{Hass1958}, $\eps_{\rm SrTiO_{3}}=6.0$~\cite{Weber1986}. Analytical curves correspond to the isotropic Maxwell-Garnett approximation with $\eps_{MG}=5+0.4\rmi $.
 }\label{fig:field}
\end{figure}

For theoretical analysis of the optical properties, we use the general semiclassical theory of light scattering~\cite{Benedek1966} as sketched below. First, the electric field $\bm E_{\rm i,\omega}(\bm r)\exp (-\rmi \omega t)$ of the plane wave incident upon the sample at the frequency $\omega$ is calculated. The field spatial distribution is strongly inhomogeneous and it is modified by the interaction with the wires. Second, the electromagnetic polarization induced in the wires by the interaction with the spin waves is determined~\cite{CardonaVII,Cottam1986}:
\begin{align}
&\bm P_{\rm s,\omega-\Omega}(\bm r)=\xi\bm m^{*}_{\Omega}(\bm r)\times \bm E_{\rm i,\omega}(\bm r)\:,&&\text{(S)}\label{eq:S}\:,\\
&\bm P_{\rm s,\omega+\Omega}(\bm r)=\xi\bm m_{\Omega}(\bm r)\times \bm E_{\rm i,\omega}(\bm r)\:,&&\text{(AS)}\label{eq:aS}\:.
\end{align}
Here, $\bm m_{\Omega}(\bm r)$ is the magnetization profile of the given spin mode with the frequency $\Omega$, and $\xi$ is the interaction constant. Equations~\eqref{eq:S}
and \eqref{eq:aS} correspond to Stokes and anti-Stokes scattering, respectively.   Depending on whether the magnon is destructed or created,  either $\bm m$ or $\bm m^{*}$ enter the expression for polarization. Clearly, the absolute values of the polarizations induced at Stokes and anti-Stokes processes can differ provided that the electromagnetic and spin waves have nonzero ellipticity.
Finally, the detected field is determined as
$
\bm E_{\rm s,\omega\pm\Omega}(\bm r)=\int\rmd^{3}r' \hat G_{\omega\pm\Omega}(\bm r,\bm r')\bm P_{\rm s,\omega\pm\Omega}(\bm r')\:,\label{eq:Green}
$
where $\hat G_{\omega\pm\Omega}(\bm r,\bm r')$ is the tensor electromagnetic Green function at the corresponding frequency.

In the experimental geometry, the incident wave is $p$-polarized, i.e. the electric field is in the $xz$ plane (see Fig.~\ref{fig:1}). The scattered wave is detected in $s$ polarization, i.e. the electric field is parallel to the $y$ axis. We have verified numerically, that  $|G_{yy}|\gg |G_{yx}|$
and $|E_{{\rm i},y}|\ll |E_{{\rm i},x}|, |E_{{\rm i},z}|$ inside the wire regions, i.e. the (linear) mixing between $s$ and $p$ polarizations inside the wires is negligible. Hence, the scattering is determined by the $y$ component of the polarizations Eq.~\eqref{eq:S} and Eq.~\eqref{eq:aS}, parallel to the detector polarization. The asymmetry between Stokes and anti-Stokes scattering can be quantified by the difference of the intensities
$|P_{\rm s,\omega-\Omega,y}|^{2}-|P_{\rm s,\omega+\Omega,y}|^{2}$ that is equal to
\begin{equation}
\Delta {I_{\rm S/AS}}
=|\xi|^{2}|\bm m_{\Omega}|^{2}|\bm E_{\rm in}|^{2}  P_{{\rm circ},y}^{(m)}P_{{\rm circ},y}^{(E)}\:,\label{eq:Delta}
\end{equation}
where we introduce the coordinate-dependent circular polarization degree of the wave $\bm e$
\begin{equation}
P_{{\rm circ},y}(\bm r)=\frac1{|\bm e|^{2}}\rmi [\bm e^{*}\times \bm e]_{y}\equiv \frac{2\Im e^{*}_{z}(\bm r)e_{x}(\bm r)}{|\bm e(\bm r)|^{2}}\:.\label{eq:P}
\end{equation}
The quantity~\eqref{eq:P} changes from $1$,  for right-circularly polarized fields, to $-1$, for left-circularly polarized and it vanishes for linear polarization. Equation~\eqref{eq:Delta} demonstrates, that the Stokes - anti-Stokes  asymmetry requires the non-zero ellipticity of both interacting waves. For the spin wave $\bm m$ polarization is dominated by the ferromagnetic resonance and is fairly close to circular. Our estimations show that it is of the order of 0.8. Another necessary ingredient for the S/AS asymmetry is the non-zero local circular polarization degree of the incident wave $P_{{\rm circ},y}^{(E)}$. Moreover, to explain the experimentally observed unexpectedly strong S/AS asymmetry it is required that it is close to the SW ellipticity, i.e. 0.8. To justify the observed inversion of the S/AS asymmetry the optical ellipticity should be reversed with respect to the case of a continuous Co film. Even though the incident electromagnetic wave is linearly ($p$)  polarized, the finite ellipticity is induced due to its refraction at the interfaces. This effect can be most simply illustrated for the case of thick continuous Co film.
The $p$-polarization vector inside the sample is equal to
$
\bm e_{p}\propto k_{z}\bm e_{x}-k_{x}\bm e_{z}\:,
$
where $k_{x}=\omega\sin\theta/c$ is the in-plane wave vector determined by the incidence angle $\theta$ and $k_{z}=\sqrt{(\omega/c)^{2}\eps_{\rm Co}-k_{x}^{2}}$. Since the permittivity of Co at the considered wavelength $\lambda=532~$nm is complex,  the $z$ component of the wave vector is complex as well, and the local electric field  is elliptically polarized with $P_{\rm circ}\approx -0.13$ at the incidence angle $\theta=20^{\circ}$. A crude Maxwell-Garnett model~\cite{sihvola1993} for the Co/CeO$_{2}$ nanowire array describes it as a slightly lossy dielectric with the averaged permittivity $\bar \eps_{\rm MG}=2(\eps_{xx}+\eps_{yy})/3+\eps_{zz}/3\approx 5+0.4\rm i$. This Maxwell-Garnett  approach yields $P_{\rm circ,y}^{(E)}\approx -0.012$, i.e. even smaller and also negative circular polarization. Both these predictions are in stark contradiction to experiment. In order to resolve this controversy we have resorted to full-wave numerical simulation of the electric field profile inside the wires and its circular polarization degree using the CST Microwave  Studio software package. The results are presented in Fig.~\ref{fig:field}:
panel (a) shows the spatial map of the circular polarization degree within the array unit cell. Figure~\ref{fig:field}(b) and Fig.~\ref{fig:field}(c) show the $z$-dependence of the  polarization at the wire center and of averaged electric field (thick/black lines).  Thin/red lines correspond to the analytical isotropic Maxwell-Garnett model. One can see, that the Maxwell-Garnett approximation well describes the distribution of the field and the polarization in the air (the region $z<0$), governed by the interference between incident and specularly reflected waves. The field decay in the sample ($z>0$) along the wire axis is captured by the Maxwell-Garnett approximation as well, see Fig.~\ref{fig:field}(c).  However, the circular polarization inside the wires is strongly different from the effective medium model. Contrary to the naive effective medium approximation, the circular polarization inside the wires has a positive sign and is quite large. Namely, $P_{\rm circ}$ oscillates around the value $+0.59$ which greatly exceeds the values both for continuous Co film ($P_{\rm circ}\approx-0.13$) and for the effective medium ($P_{\rm circ}=-0.012$). These numerical findings fully explain our experimental data.

\begin{figure}[t!]
\centering\includegraphics[width=0.45\textwidth]{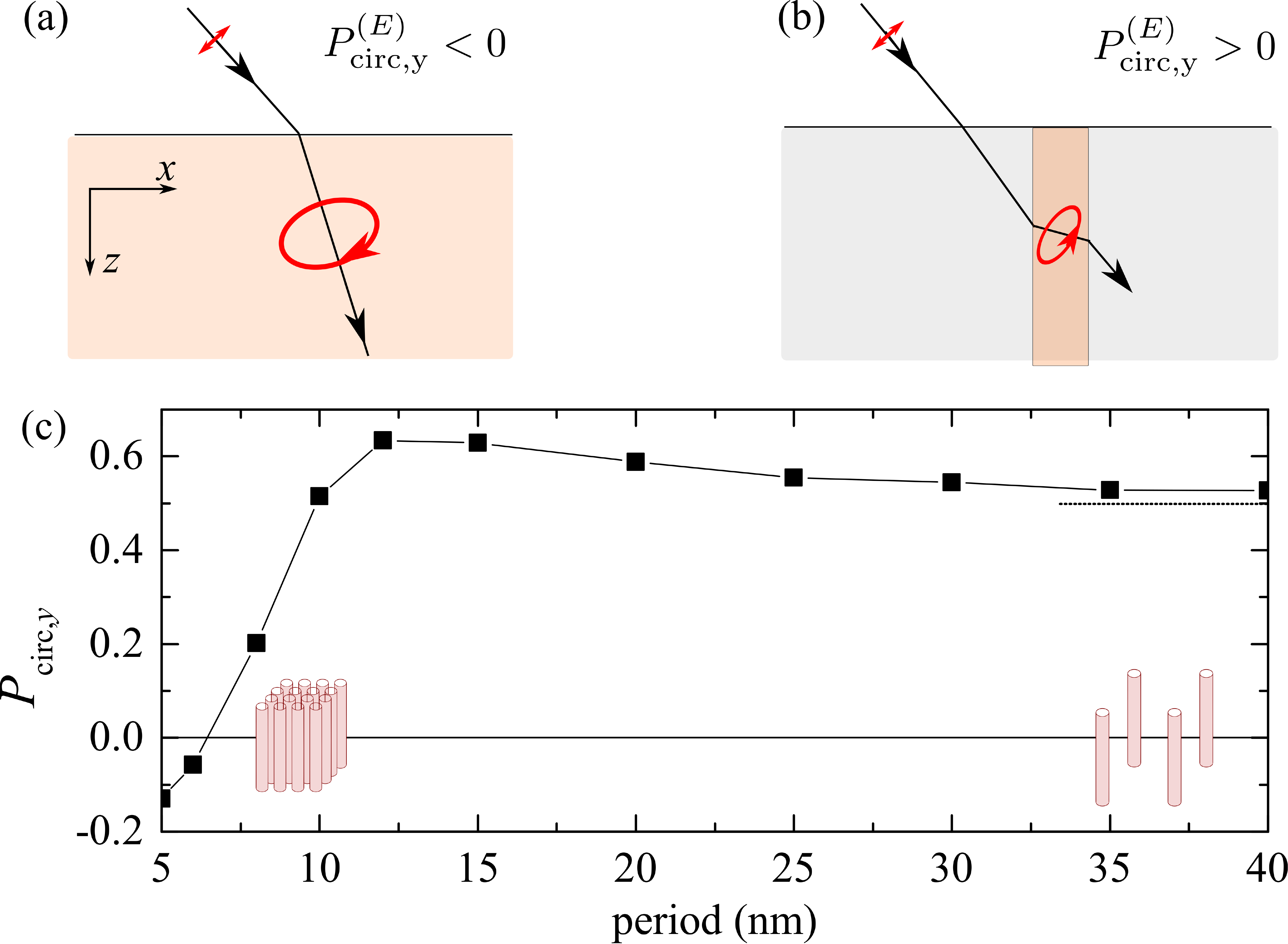}
\caption{Schematic illustration of the emergence of the  circular polarization of difference signs for the $p$-polarized plane wave, incident upon (a) positively refracting medium  and (b) wire array. (c) Dependence of averaged circular polarization inside the wires with $D=4.7$~nm on the lattice period $a$. Dotted line shows the value for the $a=400$~nm.
 }\label{fig:explanation}
\end{figure}

The polarization reversal can be qualitatively understood in the  geometrical optics approximation by examining the refraction of a plane wave at  interfaces of the structure, as shown in Fig.~\ref{fig:explanation}(a,b).  In the case of conventional positive refraction at the surface of air and lossy metal or dielectric one has $\Im k_{z}>0$, $k_{x}>0$; so that $\Im e_{p,x}>0$ and $ e_{p,z}<0$ and hence Eq.~\eqref{eq:P} yields $P_{\rm circ,y}<0$ (panel a).
%
On the other hand,  the field is actually incident upon the ultrathin wires from the side, rather than from the top face (panel b). As a result, the imaginary parts of the components of the field polarization vector inside the wires are  swapped ($e_{p,x}>0$ and $ \Im e_{p,z}<0$ ), and the sign of the circular polarization is reversed with respect to the continuous film.
The crossover between the single wire and continuous film limit is illustrated by the polarization dependence on the lattice period shown in Fig.~\ref{fig:explanation}(c). The polarization is positive for large periods in agreement with the geometric optics prediction  (dotted line). For smaller periods the polarization is increased due to the collective effects and it has a flat maximum at the spacings $a\sim 10\div 15$~nm, close to the experimental data. For even smaller periods the polarization decreases and it changes the sign as a dense wire array approaches  the continuous film limit.

A good quantitative agreement between our theory and experiment is illustrated in Fig.~\ref{fig:spectra}(e). In this figure we trace a ratio of the intensities of anti-Stokes and Stokes BLS lines as a function of the angle of incidence, both experimental and theoretical values obtained using Eq.~\eqref{eq:Delta}. In  theoretical analysis we use the values of optical ellipticity obtained from our CST numerical simulations (see Fig.~\ref{fig:field}) while the same parameter for the SW is estimated using the analytical solution for Kittel modes on an individual ferromagnetic wire. To facilitate understanding the angular evolution of the ellipticity of the optical polarization  is given in the inset of Fig.~\ref{fig:spectra}(e).

In conclusion, by employing the Brillouin light scattering tools we have observed the effect of anomalous polarization conversion in arrays of ultrathin Co nanowires with the nanowire diameter below 5 nm manifested in the pronounced reversed Stokes/anti-Stokes scattering asymmetry. We have explained this effect by unexpected circular polarization of light induced within the nanowires. In particular, the circular polarization has the opposite sign being much larger in the absolute value than that for the continuous films or thicker nanowires. This finding opens a great potential of seemingly simple nanowire arrays employed for manipulating light at the nanoscale. At the same time, our results suggest that the Brillouin spectroscopy, traditionally employed as a probe of magnon states, is an extremely sensitive technique for studying polarization-resolved landscapes of local electric fields in nanostructures.

\begin{acknowledgements}
The authors acknowledge useful discussions with B. Jusserand and E.L. Ivchenko.We thank D. Demaille for TEM microscopy and J.-M. Guigner, IMPMC, CNRS-UPMC, for access to the TEM facilities. This work was supported by the Government of Russian Federation (Grant 074-U01), the Dynasty Foundation (Russia), and the Australian Research Council. ANP acknowledges a support by RFBR and the Deutsche Forschungsgemeinschaft under the International Collaborative Research Center TRR 160. A part of this work was supported by ANR (ANR-2011-BS04-007).
\end{acknowledgements}

%

\newpage

\section*{Supplementary Information}
\renewcommand{\thefigure}{S\arabic{figure}}
\setcounter{figure}{0}

Here we present the details of the derivation of the results for the spin mode frequency, that have been used  in order to identify the dominantly scattering spin mode.
In order to obtain the spin mode frequency as a function of the wave vector $q$ we average the Landau-Lifshitz equation of motion together with the Maxwell magnetic flux conservation over the cross-section of the cylinder, thus generalizing, for the cylindrical symmetry, the approach, originally proposed for thin films by  Stamps and Hillebrands \cite{Stamps1991} and revisited later in Ref.~\cite{Zighem2007}.
The applicability of this technique, as mentioned before, is limited to the case where the transversal distribution of the dynamic magnetization is close to uniform, which corresponds to the Kittel mode we are interested in. This corresponds, in the Damon-Eshbach configuration, to the lowest fundamental mode, referred to in this article as ``Kittel mode''~\cite{KittelIntro}. Below are sketched major steps in this calculation. A detailed account will be published elsewhere.

{\bf A. } The aim of this part is to derive the dynamic magnetic field averaged over the section. 
In this section we use the coordinate system where the wire axis is directed along $z$.
In the frame of the quasi static approximation, the dynamic magnetic field $\bm h$ is obtained as the gradient of a potential 
\[
\bm h=\bm\nabla\Phi\:.
\]
We assume a propagation along the nanowire axis $z$, $\Phi(x,y,z)=f(x,y)\exp(\rmi qz)$\:.
As the probed mode are quasiuniform across the section, we consider the dynamic magnetic field averaged over the section:
\[\langle h_{x}\rangle=\left\langle\frac{\partial \Phi}{\partial x}\right\rangle\:,\langle
 h_{y}\rangle=\left\langle\frac{\partial \Phi}{\partial y}\right\rangle,\langle h_{z}\rangle=\rmi q \langle\Phi\rangle\:.
\]
Next, we use the cylindrical coordinate system with $\phi$ being the azimuthal angle in the $xy$ plane and $\rho=\sqrt{x^{2}+y^{2}}$ being the two-dimensional radius vector. Using the Stokes theorem one can express the functions $\langle \partial\Phi/\partial{x}\rangle$,  $\langle \partial\Phi/\partial{y}\rangle$,  and $\rmi q \langle\Phi\rangle$  via the values of  $f(\rho,\phi)$ evaluated at the nanowire surface where $\rho=R$. The function $f(R,\phi)$ can be expanded as 
\[
f(R,\phi)=f_{0}+f_{1}\cos\phi+f_{2}\sin\phi\:.
\]
The coefficients $f_{0}$, $f_{1}$, $f_{2}$ are derived from the boundary conditions:  
\[
f_{0}\approx\langle f\rangle,\quad f_{1}\approx  -2\pi R\langle m_{x}\rangle ,\quad  f_{2}\approx  -2\pi R\langle m_{y}\rangle\:.
\]
Thus one obtains 
\[
\langle h_{x}\rangle=-2\pi\langle m_{x}\rangle,\quad
\langle h_{y}\rangle=-2\pi\langle m_{y}\rangle\:,
\]
\[
\langle h_{z}\rangle=-\frac{4\pi\langle m_{z}\rangle}{1-\frac{2}{qR}\frac{K_{0}'(qR)}{K_{0}(qR)}}\:.
\]
 where $K_{0}$ is a modified Bessel function. 

{\bf B. } In this part, the frequency from the averaged equation of motion is derived assuming an effective anisotropy energy $\pi M_{x}^{2}+\pi M_{y}^{2}-KM_{z}^{2}/M^{2}$, where $K$ contains the magneto-crystalline contribution and the dipolar coupling. 

{\bf B.1} First, we consider the case when the applied field is not saturating. The effective field reads $H_{y} =H-2\pi M_{y}$ , $H_{z} =2KM_{z} /M^{2}$. 
The equilibrium condition reads $H = (2\pi +2K/M^{2})M_{y}$. The averaged equations of motion yield
\begin{multline*}
\rmi (\Omega/\gamma) \langle m_{x}\rangle= M_{y} [ \langle h_{z}\rangle - 2Aq^{2}\langle m_{z}\rangle/M^{2}+2K
\langle m_{z}\rangle
/M^{2}]\\-M_{z}[ \langle h_{y}\rangle - 2Aq^{2}\langle m_{y}\rangle/M^{2}] +\langle m_{y}\rangle H_{z}-\langle m_{z}\rangle H_{y}\:.
\end{multline*}
\[
\rmi (\Omega/\gamma)\langle m_{y}\rangle= M_{z} [ \langle h_{x}\rangle - 2Aq^{2}\langle m_{x}\rangle/M^{2}] - \langle m_{x}\rangle H_{z}\:,
\]
\[
\rmi (\Omega/\gamma)\langle m_{z}\rangle= -M_{y} [ \langle h_{x}\rangle - 2Aq^{2}\langle m_{x}\rangle/M^{2}] + \langle m_{x}\rangle H_{y}\:.
\]
Replacing the averaged field components by the expressions derived in the previous part, one obtains
\begin{multline*}
\left(\frac{\Omega}{\gamma}\right)^{2}=\Bigl[ M_{y}\Bigl(\frac{4\pi}{1-\frac{2}{qR}\frac{K_{0}'(qR)}{K_{0}(qR)}}\Bigr)
+\frac{2Aq^{2}}{M^{2}}-\frac{2K}{M^{2}}
+H_{y}\Bigr]\\\times\Bigl[ M_{y}\left(2\pi+\frac{2q^{2}A}{M^{2}}\right) +H_{y}\Bigr]\\+
\left( M_{z}\left(2\pi+\frac{2q^{2}A}{M^{2}}\right)+H_{z}\right)^{2}\:.
\end{multline*}

{\bf B.2} Second, we consider the case the applied field is saturating. Using the method presented before, one obtains
\begin{multline*}
\left(\frac{\Omega}{\gamma}\right)^{2}=\left(H+\frac{2q^{2}A}{M}\right)
\Bigl(\frac{4\pi}{1-\frac{2}{qR}\frac{K_{0}'(qR)}{K_{0}(qR)}}\\+\frac{2Aq^{2}}{M}-\frac{2K}{M}+H-2 \pi M\Bigr)\:.
\end{multline*}

{\bf C.}
In order to identify the origin of the dominant scattering mode we have measured the dependence of the Brillouin shift frequency $\Omega$ versus the applied magnetic field $H$; the results are presented in Fig.~\ref{fig:field}. Interestingly, the $\Omega(H)$ curve demonstrates a pronounced minimum:  the frequency decreases for low values of $H$ while displaying a clearly seen growth after passing the critical point. This characteristic softening is related to the saturation of the static magnetization. Red curve shows the calculation according to the theory above. The effective value of the wave vector $q$ of the spin wave, contributing to the Brillouin scattering has been deduced by fitting the experimental results. According to these calculations the wavelength of the ``scattering magnon'' is $2\pi/q\sim 100$~nm. 

\begin{figure}
\centering\includegraphics[width=0.45\textwidth]{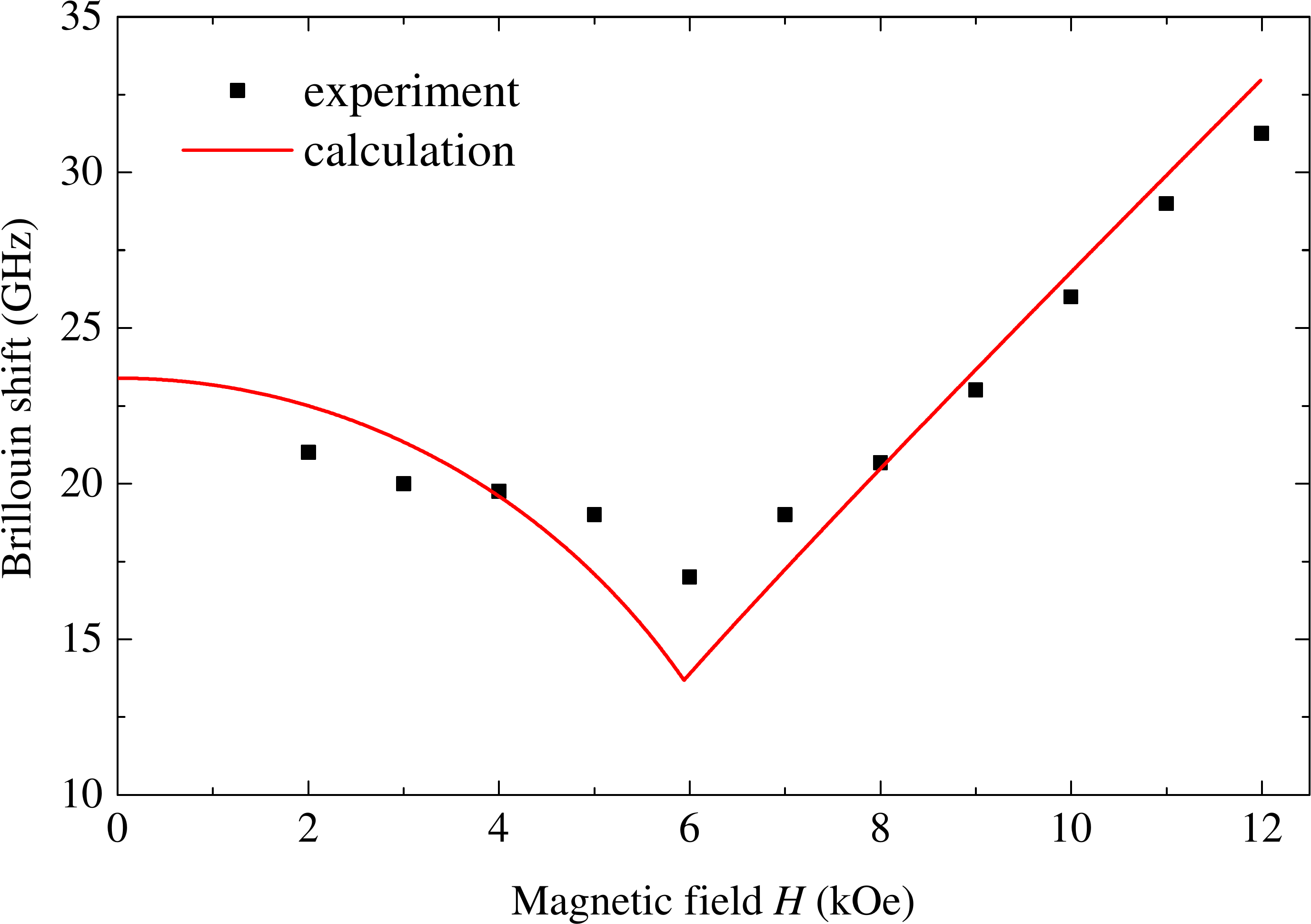}
\caption{Dependence of the spin mode frequency on the applied field strength. Squares are experimental results, the line has been calculated with the following set of parameters:
 $M$ = 1400~emu/cm$^{3}$, A = 1.3$\times 10^{-6}$erg/cm, 
  $\gamma/(2\pi)= 3~$GHz/kOe, $R=2.4$~nm, 
  $K = -1.8\times 10^{6}$~erg/cm$^{3}$. }\label{fig:field}
\end{figure}

\end{document}